# Analogue-dynamical Prediction of Numerical Model Errors in the Mid-Lower Reaches of the Yangtze River

Qiguang Wang    Aixia Feng
College of Atmospheric Sciences of Lanzhou University,
Lanzhou, China

Guolin Feng   Zhihai Zheng
Laboratory for Climate Studies
NCC of China Meteorological Administration
Beijing, China

*Abstract*—**A new prediction error correction scheme based on 74 circulation characteristics data provided by Weather Diagnostic Forecasting Division of National Climate Center, which is designed to develop the Operational Numerical Forecast Model (ONFM) of the National Climate Center of China, and the skill level of the precipitation prediction for rainy season in the mid-lower reaches (MLR) of the Yangtze River by ONFM is obviously raised. The approach use principal component(PC) analysis to prediction error of ONFM. And we used different factors to correct the different PCs of the error of precipitation field. The comparative study results indicate that the effectiveness of the new analogue error correction (AEC) scheme is better than system error correction (SEC) scheme.**

## I. Introduction

Seasonal forecast for China has been reviewed as scientific challenges, especially for the region of the Mid-Lower Reaches of the Yangtze River, and the skill level of the summer precipitation prediction was not always high [1-3]. In order to improve the summer precipitation prediction of the region, a series of researches were done in a brand new perspective, such as techniques of ensemble prediction, physical model forecasting and processing after the model and so on[4-8]. Both of the statistics techniques and dynamical schemes of seasonal prediction were advanced and improved with the accumulated of the data and the improved model[9-11]. For instance, a new scheme was proposed based on regional inter-annual incrimination, and made a successful simulation on the trend of the precipitation in the region. Combination of multi-model data and downscaling technique, the precipitation forecast has made some progress in Chinese regions. Besides, combination of dynamics and statistics is one of effective ways to advance the short-term climate prediction. The theoretical feasibility of using the historical data in the numerical prediction through statistical and dynamical combination was demonstrated by Chou[7]. The air motion had self-memory, and derived self-memory equations of the multi-time observed data[12-16]. And they improved the short-term climate prediction effectively. However, generally speaking, the current operational prediction skill still does not meet social demand, and remains to be further improved.

The factors affecting the summer precipitation in the Mid-Lower Reaches of the Yangtze River are intricate. Sometimes, the precipitation is the result of the co-action of the factors, and sometimes is the contribution of part or single factor. This paper predicted the error of the model based on historical analogue information through decomposition and compression the dimension of the error field by EOF method. We proposed a new method of model error prediction with a matched spatial and temporal scale and mainly for the first three principal components. We applied the method to regress the summer precipitation in the region with latitudes from 27.5°N to 32.5°N and longitudes from 110°E to 122.5°E.

## II. Data and Method

### A. Data

The data used in this paper is the rainy season rainfall (RSR) forecast data set of the atmosphere-ocean coupled ONFM in the period 1983-2009, which was generated by the ONFM without prediction error correction. The yearly prediction error fields of RSR were calculated from the hindcast precipitation fields of ONFM and the RSR data of the CPC Merged Analysis of Precipitation (CMAP, as the observation) in the 27-yr, and the climate mean state of RSR was calculated from the CMAP data of 1983-2002. Analogues were selected from the 74 characteristics data from the National Climate Prediction Center Climate System Diagnostic Division.

### B. The basic principle of analogue-dynamical prediction and PC analysis

In general, the numerical prediction model is of as initial value problem of partial differential equations, which are expressed as follows:

$$\begin{cases} \dfrac{\partial y}{\partial t} + L(y) = 0 \\ y(x, t_0) = y_0(x) \end{cases} \quad (1)$$

Where $y(x, t_0)$ is the model state vector to be predicted, $x$ and $t$, respectively, the vector in spatial coordinate and time, $L$ is the differential operator of $y$, which is corresponding to a real numerical model, 0 the initial time, $y_0$ the initial value.

Then the exact model that real atmosphere satisfies can be written as:

$$\begin{cases} \frac{\partial \tilde{y}}{\partial t} + L(\tilde{y}) = E(\tilde{y}) \\ \tilde{y}(x,0) = \tilde{y}_0(x) \end{cases} \quad (2)$$

Where $E$ is the error operator expresses the errors of real numerical model. Combination Eq.2 and Eq.1 [14], the outcomes of the model forecasting is

$$\hat{P}(y_0) = P(y_0) + \hat{P}(\tilde{y}_j) - P(\tilde{y}_j) \quad (3)$$

Where $\hat{P}(y_0)$ is the forecasting through the error analogue estimate, $P(y_0)$ is the output of $y_0$ by the model, $\hat{P}(\tilde{y}_j)$ is the in situ of the historical analogue and $P(\tilde{y}_j)$ is the forecasting of the historical analogue initial values. The essential of the equation is to estimate the current forecasting errors through historical analogue errors information. And by that we could reduce the model error and turn the dynamical forecasting into estimate the forecasting errors.

Principal component analysis has become a widely used tool in climate research. It also known as empirical orthogonal function (EOF) analysis. Principal component analysis can reduce the number of degrees of the data without losing much of the information. EOF analysis on the element field $j_{m \times n}$ can be expressed as:

$$j_{m \times n} = V_{m \times n} T_{m \times n} \quad (4)$$

Here $m$ and $n$ is the space and time dimensions respectively. $V_{m \times n}$ is the normalized orthogonal basis, $T_{m \times n}$ is the time coefficients (PCs).

III. FORECASTING THE MODEL ERRORS WITH ANALOGUE-DYNAMICAL PREDICTION

The error correction results were displayed in Figure.1 by applying three indexes analogue error correction. And the three factors are the index of global averaged land and ocean, North America subtropical high intensity and West Pacific subtropical high area. The scheme is that normalized the three index time series respectively, extracted the principal components occupying 80% explained variance by EOF, determined the four rainfall error field years through the time coefficients of the principal components, and finally the cross-examined results of 27-yr analogue error correction were obtained. The forecasting analogue error field is the average model error fields of the four analogue years just as the formalism-(5).

$$AI = \frac{\sum_{i=1}^{k} w_i (j'_{ij} - j'_{ik})^2}{\sum_{i=1}^{k} w_i} \quad (5)$$

Where $AI$ is the Euclidean distance between the time series of the principal components, $j$ and $k$ represent filed of two different times, $w_i$ is the weighted coefficient. The smaller the Euclidean distance is, the similar the fields are. Here the impact and connections between indexes could be removed by EOF. Meanwhile the nonlinear error increase could be excluded and the dimensions could be reduced as well. The average 27-yr Anomaly Correlation Coefficient(ACC) is up to 0.29 by the above scheme, but that of the systematic (sys) correction is only -0.01 much smaller than the analogue correction. Therefore the analogue (ana) correction method is remarkable.

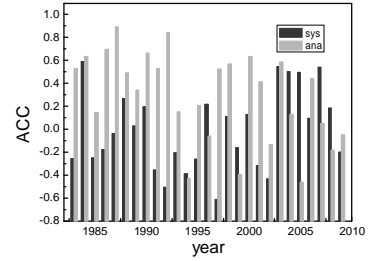

Figure 1. The 27-yr ACC of three indexes analogue and systematic correction

IV. THE FORECASTING BASED ON COMPRESSION DIMENSIONS OF THE ERROR FIELD OF THE MODEL

A. Forecasting the Anomaly of Errors with PCs

The Figure 2(a) depicted the ACC of different principle component in the regions researched after EOF decomposition about the anomaly of errors. It can be found that the accumulative ACC between the forecasting and the in situ field is up to 0.12 much higher than the system error correction -0.01 for the first three principal components whose accumulative explained variance are 80%. Then the ACC is in a almost steady state with the number of principal components increasing. The first principal component and other components higher than four have little contribution to the ACC, but the second and third component contribute more to the ACC lead it up to 0.10, just as Figure 2(b) illustrated. So we can take the first three principal components for the processes of the analogue error correction. It has virtues such as reducing the calculation and filtering out part of the impact from random errors.

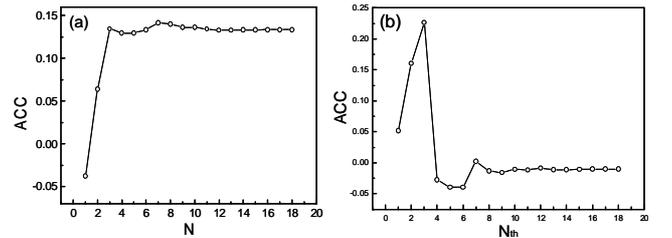

Figure 2. ACC of analogue error correction through cross-examination under different numbers of principal component: (a) accumulative principle component, and (b) single principle component

Some researches demonstrate that the precipitation of the Mid-Lower reaches of the Yangtze River has characteristics of quasi-two periodic and decadal oscillation. Therefore the forecasting must object to different time scales. The next section would center the prediction of the first three components of different time scales.

B. Cross-validation

The Pacific Decadal Oscillation (PDO) has close connection to the summer precipitation of East China. And the Quasi-Biennial Oscillation(QBO) ,the Northern Boundary of the Subtropical High(NBSH) of the atmospheric air also has strong connection with the rainfall of the rainy season in our country. It is also true for the subtropical high intensity index of the South Sea in the pre-December. Therefore we used the above three factors to represent different time scale factors and make predication of the first three factors for the precipitation field, the outcomes just as Figure 3 shown.

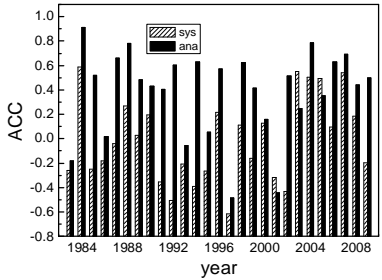

Figure 3. ACC of analogue error correction through cross-examination under different numbers of principle component: the twenty seven years cross-examined ACC

As Figure 3 illustrated the forecasting can be improved further by using different time scale factors. And the ACC of the twenty seven years was from -0.01 by system error correction up to 0.38. It showed the potential operational values of our schemes.

*C. The flow of the scheme*

Based on the above numerical experimentation result, the anomaly filed of the mid-lower Yangtze River was decomposed by EOF to correct the first three components by the analogue correction using the pre-indexes. And the chart of the scheme is just as the Figure 4 shown.

1. The forecasting error filed year by year was obtained based on the CMAP rainfall data and hindcast of the ONPM.

2．The first three modes are the objections to be corrected by the analogue error correction through the EOF to decompose the anomaly error filed of the ONPM.

3. The pre-factors of the forecasting year in rainy season are the 74 indexes in January of the same year and that in from February to December of the year before. The first three modes were corrected by analogue correction through singular cross-examination, and the other modes were corrected by systematic correction. The results were compared with that of the CMAP. The rank of average ACC from 1983 to 2008 could be obtained through singular analogue correction.

4. For the first three components, the principal factors are those with the first rank of ACC. Then other indexes are added based on their ranks to obtain the optimal factor combination through cross-examination.

5. The error filed of the forecasting year of the ONPM was decomposed by EOF. For the first three components of the EOF, they were corrected by analogue error correction. Other components were instead of climatic state.

6. Finally, the forecasting result of the rainy season rainfall was obtained. That was the result of the forecasting model plus the analogue correction error.

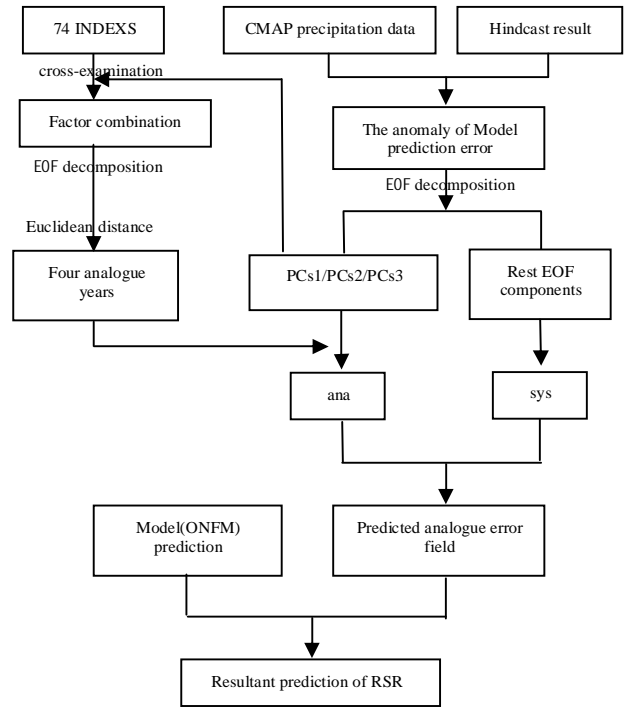

Figure 4. Flow chart for AEC-PCs scheme

*D. Independent sample hindcasts for 2003 to 2009*

The hindcast results of rainy season in mid-lower Yangtze River from 2005-2009 were obtained through independent sample experiment just as the Figure 5 shown. The scheme used was ONPM to make sure using efficient information of the history.

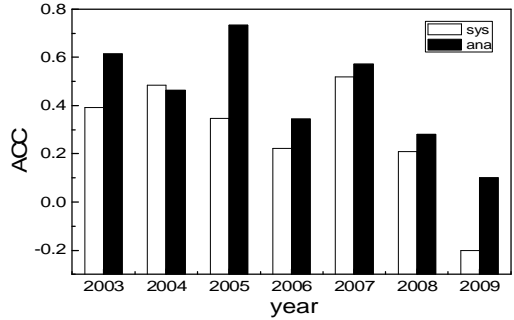

Figure 5. The ACC of the 2003-2009 independent sample hindcasts

The results of our study show that the analogue error correction is better than the systematic correction. The details are just as following. For the first three components of error anomaly fields from 2003 to 2009, the optimal factor combinations were going to obtain for the analogue error correction. The other steps were just the scheme chart. The ACC of the studied seven year is 0.61, 0.46, 0.73, 0.34, 0.57, 0.28, and 0.1 respectively. The systematic correction ACC of 2009 is -0.2, though the average ACC of the seven year is 0.28. That means the instability of the systematic correction. Meanwhile the average ACC of the seven year by analogue correction is 0.44, and the increase of the ACC is up to 0.16 .

## V. Summary and discussion

The errors of the model were predicted based on historical useful information through decomposition and compression the dimension of the error field, and division the scaled quantity out of the field. The first three principal components were predicted based on the combination of the matched spatiotemporal pre-factors respectively. As for the small-scale components, the probability of the climate mean state was given. The Anomaly Correlation Coefficient of the seven years was from 0.28 by system error correction up to 0.44, and reached a relative high level.

## Acknowledgment

Funding was obtained from Public Service Sectors Special Research under GYHY No. 200806005, National Natural Science Foundation of China under Grant No.41005041, and National Science and Technology Support program under Grant No. 2007BAC29B01.